\documentstyle[psfig,tighten,preprint,aps]{revtex}

\begin{document}
\draft
\preprint{\vbox{\it Submitted to Phys. Rev. D \hfill\rm TRI-PP-95-62}}

\title{Abelian dominance of chiral symmetry breaking in lattice QCD}
\author{Frank X. Lee, R.M. Woloshyn}
\address{TRIUMF, 4004 Wesbrook Mall,
Vancouver, British Columbia, Canada V6T 2A3}
\author{Howard D. Trottier}
\address{Department of Physics, Simon Fraser University, Burnaby,
British Columbia, Canada V5A 1S6}
\date{\today}
\maketitle

\begin{abstract}
Calculations of the chiral condensate
$\langle \bar{\chi} \chi \rangle$ on the lattice
using staggered fermions and the Lanczos algorithm are presented.
Four gauge fields are considered:
the quenched non-Abelian field, an Abelian projected field, and
monopole and photon fields further decomposed from the Abelian field.
Abelian projection is performed in maximal Abelian
gauge and in Polyakov gauge.
The results show that monopoles in maximal Abelian gauge
largely reproduce the chiral condensate values of
the full non-Abelian theory, in both SU(2) and SU(3) color.
\end{abstract}
\vspace{1cm}
\pacs{PACS numbers: 11.15Ha,11.30.Rd}

\section{Introduction}
\narrowtext

Since the Abelian monopole mechanism for
confinement in QCD was first proposed \cite{hoof1,hoof2},
there have been extensive studies in the pure gauge sector
of lattice theories \cite{suzuki,kron,kron1,brand,deb,yee,poulis}.
The effect of Abelian projection in the quark sector was
studied in \cite{rmw}, where chiral symmetry breaking
and meson correlators were analyzed.
Further studies of the role of Abelian monopoles in
chiral symmetry breaking at finite temperature \cite{miya},
and in hadron spectroscopy \cite{suzuki1}, have been made.
These works have provided evidence in support of
Abelian monopoles as the mechanism of chiral symmetry breaking
in QCD.

In the present work, a systematic study of the
role of Abelian projection and monopoles in chiral
symmetry breaking is carried out in the confined phase at zero temperature.
The chiral condensate is computed using four sets of
gauge fields: the quenched non-Abelian field, an Abelian projected
field, and monopole and photon fields further decomposed from
the Abelian projected field. Abelian projection is performed in
maximal Abelian gauge and in Polyakov gauge.
We employ a different technique to compute the
chiral condensate than was used in Ref. \cite{rmw},
where an extrapolation to the chiral limit was made from nonzero quark
mass: here we use the spectral representation of the chiral condensate
along with the Lanczos algorithm \cite{lanc}, which allows
calculations to be done
directly at zero mass. Most of the results presented here are for
SU(2) color. Some results of a first study of chiral symmetry breaking
in the Abelian projection of SU(3) gauge theory are also given.

The calculational method is described in Sec. \ref{met}
and results are presented in Sec. \ref{res}.
Using the Lanczos method at zero quark mass we find reasonable
quantitative agreement between the chiral condensate calculated with the
monopole part of the Abelian projected field and the quenched non-Abelian
field for the SU(2) theory. Our present SU(3) simulation is not yet good
enough to draw the same quantitative conclusion, but qualitatively, the
same pattern emerges as seen in the SU(2) calculation. This provides some
evidence that the Abelian dominance idea can be extended to the SU(3) theory.

Some of our results were presented in
summary form in Ref. \cite{lat95}.

\section{Method}
\label{met}

\subsection{Chiral Condensate on the lattice}

Chiral symmetry breaking is studied using staggered fermions,
with the action
\begin{eqnarray}
   S_f & = & \frac{1}{2} \sum_{x,\mu} \eta_\mu(x)
   \left[ \overline{\chi}(x) \, U_\mu(x)
   \,\chi(x+\hat{\mu}) \,\, -\overline{\chi}(x+\hat{\mu})
   U^{\dagger}_\mu(x) \, \chi(x) \right]
   + \sum_x \, m \overline{\chi}(x) \chi(x),
\nonumber \\
   & \equiv & \overline{\chi} \,\, {\cal M}(\{U\}) \, \chi  ,
\label{Sf}
\end{eqnarray}
where $\overline{\chi}, \chi$ are single-component fermion fields,
$\eta_\mu(x)$ is the staggered fermion phase \cite{kawo},
$m$ is the mass in lattice units and the $U's$ are gauge field links.

The chiral symmetry order parameter is calculated from the
inverse of the fermion matrix ${\cal M}$ of Eq. (\ref{Sf})
\begin{eqnarray}
   \langle \overline{\chi} \chi(m,V) \rangle
   = \frac{1}{V} \, \langle \mbox{Tr} \, {\cal M}^{-1}(\{U\}) \rangle ,
\end{eqnarray}
where V is the lattice volume and the angle brackets
denote the gauge field configuration average.
This can be rewritten in a spectral representation~\cite{bar}
\begin{equation}
\langle \bar{\chi} \chi(m,V) \rangle ={-1\over V}
\sum_{n=1}^N {1\over i\lambda_n +m}
={-1\over V}\sum_{\lambda_n\geq 0}{2m\over \lambda_n^2 +m^2},
\label{spec}
\end{equation}
where $i\lambda_n$'s are the eigenvalues of the zero-mass fermion
matrix (Dirac operator).
The eigenvalues appear in complex conjugate pairs so
only the positive half of the spectrum needs to be considered as 
in the second part of Eq.~(\ref{spec}).

To correctly probe the physics of spontaneous chiral symmetry breaking,
one should attempt to work in the limit of zero quark mass and
infinite volume. The chiral limit $m\rightarrow 0$  should be taken
after $V\rightarrow \infty$
\begin{equation}
\langle \bar{\chi} \chi \rangle=
- \lim_{m\rightarrow 0} \lim_{V\rightarrow \infty}
{1\over V}\sum_{\lambda_n\geq 0}{2m\over \lambda_n^2 +m^2}
= -\lim_{m\rightarrow 0} \int_0^\infty
  d\lambda{2m \rho(\lambda) \over \lambda^2 +m^2}
=-\pi\rho(0),
\label{rho}
\end{equation}
where the spectral density function
$\rho(\lambda)={1\over V}{dn/ d\lambda}$ is
normalized to $\int_0^\infty d\lambda\rho(\lambda)=N_c$,
the number of colors. Eq.~(\ref{rho})
relates chiral symmetry breaking to the small modes in
the  eigenvalue spectrum.
So the task is reduced to finding the small eigenvalues of 
the fermion matrix at zero mass, rather than the entire spectrum.
This can be done using the well-established Lanczos algorithm~\cite{lanc}.

\subsection{Abelian Projection on the Lattice}
\narrowtext

The lattice formulation of Abelian projection was developed
in~\cite{kron,kron1}. The idea is to fix the gauge of a SU(N)
theory so that a residual $U(1)^{N-1}$ gauge symmetry remains.
The Abelian degrees of freedom are extracted by a subsequent
projection $ U(x,\mu)=c(x,\mu)\;u(x,\mu)$,
where $u(x,\mu)$ is the (diagonal) Abelian-projected field and
$c(x,\mu)$ the non-diagonal matter field.
In general the gauge condition can be
realized by diagonalizing some adjoint operator ${\cal R}$:
\begin{equation}
   G(x){\cal R}(x)G^{-1}(x)=\mbox{diagonal}.
\label{diag}
\end{equation}

\widetext
Several gauge conditions have been studied
and it has been found that the so-called maximal
Abelian gauge~\cite{kron1} most readily captures the
long-distance features of confinement (the relevance
of other gauges to long-distance physics has been
considered in \cite{poulis}). In SU(2) this gauge
is realized by maximizing the quantity
\begin{equation}
 R=\sum_{x,\mu} \mbox{tr} \left[
\sigma_3 \tilde{U}(x,\mu)\sigma_3 \tilde{U}^\dagger(x,\mu)\right]
\end{equation}
where $\tilde{U}(x,\mu)=G(x)U(x,\mu)G^{-1}(x+\mu)$.
We maximize $R$ by an iterative procedure.
We find a gauge transformation $G(x)$ that maximizes $R$ locally,
at given site $x$, by keeping $G(x+\mu)$ at neighboring sites
fixed. Using the fact $\sigma_3 G = G^\dagger \sigma_3$
(with the exception of diagonal gauge transformations)
leads to an analytical solution for the local maximization of $R$
\begin{equation}
   G^2(x)=F^{-1}(x),
\end{equation}
where
\begin{equation}
F(x)=\sum_{\mu=1}^4\left[U(x,\mu)\sigma_3U^\dagger(x,\mu)\sigma_3
+U^\dagger(x-\mu,\mu)\sigma_3U(x-\mu,\mu)\sigma_3 \right] .
\end{equation}
The global maximum of $R$ is approached
iteratively by repeatedly sweeping through the lattice, until $G(x)$
is sufficiently close to the identity:
\begin{equation}
   \mbox{max} \left\{ 1 - \frac12 \mbox{Tr}G(x) \right\}
   \leq \delta \ll 1 , \quad \delta \sim 10^{-6}.
\label{delta}
\end{equation}

Maximal Abelian gauge in SU(3) is implemented by maximizing
\begin{equation}
R=\sum_{x,\mu}\left[ |\tilde{U}_{11}(x,\mu)|^2
+|\tilde{U}_{22}(x,\mu)|^2+|\tilde{U}_{33}(x,\mu)|^2 \right].
\label{r3}
\end{equation}
The maximization is done by going through the three SU(2) subgroups
of SU(3). The gauge transformation within the first subgroup
can be parameterized as
\begin{equation}
G_1(x)=\left(\begin{array}{ccc}
g_0 & g_2+ig_1 & 0 \\
-g_2+ig_1 & g_0 & 0 \\
0 & 0 & 1 \end{array} \right)
\end{equation}
with the constraint $g_0^2+g_1^2+g_2^2=1$. 
$R$ is maximized locally at $x$, again by holding
$G_1$ at neighboring sites fixed. Under this local transformation,
$R$ can be expressed as
\begin{equation}
R=\sum_{x}\left[c_1g_0^2+c_2g_1^2+c_3g_2^2+c_4g_0g_1+c_5g_0g_2 \right]
\end{equation}
where the sum over $\mu$ is implicit in the coefficients $c_i$.
The solution is obtained
numerically, and this is facilitated by
introducing spherical coordinates on the unit sphere:
$g_2=\sin\theta\cos\phi$, $g_1=\sin\theta\sin\phi$, $g_0=\cos\theta$.
This procedure is repeated at a given site for the second and the
third subgroups.
The global maximum of R is approached by repeatedly
sweeping through the entire lattice until a stopping
criterion similar to Eq. (\ref{delta}) is satisfied.

Once the gauge fixing is done, the Abelian-projected
links are extracted. In SU(2) gauge theory, where the links
can be parameterized by
$\tilde{U}(x,\mu)=r_0+i\vec\sigma \cdot \vec r$,
the Abelian projected links $u$ are given by
\begin{equation}
   u=\mbox{diag}(e^{i\phi},e^{-i\phi}),
   \quad \phi=\tan^{-1}(r_3/r_0).
\end{equation}
In SU(3), the Abelian configurations are extracted according to
\begin{equation}
   u=\mbox{diag}(u_1,u_2,u_3),
   \quad
   u_k=\exp{\left(i\,\arg{\tilde{U}_{kk}}-\frac{i}{3}\varphi\right)},
\label{uuu}
\end{equation}
where
\begin{equation}
   \varphi=\left(\sum_k\arg{\tilde{U}_{kk}}\right)\;\;\mbox{mod}\;
   2\pi\subset (-\pi,\pi].
\end{equation}
The three phase factors are constrained by $u_1u_2u_3=1$
so that only two of them are independent.

\subsection{Monopole Decomposition}

It is well-known that there exist monopoles in a compact
U(1) field. The Abelian-projected link
$u(x,\mu) = \exp(i \phi_\mu(x))$ can be resolved into a
component due to monopoles by considering
the Abelian field strength $\phi_{\mu\nu}(x)$,
defined from plaquette phases in the usual way
\begin{equation}
   \phi_{\mu\nu}(x) = \partial_\mu \phi_\nu(x)
                    - \partial_\nu \phi_\mu(x) ,
\end{equation}
where $\partial_\mu f(x) = f(x+\hat\mu) - f(x)$.
The flux due to an integer-valued monopole string
$\tilde m_{\mu\nu}$ is identified from the field strength
according to \cite{DeGrand}
\begin{equation}
   \phi_{\mu\nu}(x) = \phi'_{\mu\nu}(x)
                    + 2 \pi \tilde m_{\mu\nu}(x) ,
\end{equation}
where $\phi'_{\mu\nu} \in (-\pi, \pi]$.
The vector potential $\phi^{\rm mon}_\mu(x)$ generated
by the monopoles is therefore given (in Lorentz gauge) by
\begin{equation}
   \phi^{\rm mon}_\nu(x) = -2 \pi \sum_y D(x-y)
             \partial'_\mu \tilde m_{\mu\nu}(y) ,
\end{equation}
where the lattice photon propagator satisfies
$-\partial'_\mu \partial_\mu D(x) = \delta_{x,0}$,
with $\partial'_\mu f(x) = f(x) - f(x-\hat\mu)$ \cite{smit}.
The ``photon'' field $\phi'_\mu$ is identified with the
difference $\phi_\mu(x) - \phi^{\rm mon}_\mu(x)$.

\section{Numerical results}
\label{res}

The Lanczos algorithm~\cite{lanc} is carried out in double precision
to find the eigenvalues of the fermion matrix.
To test our implementation of the algorithm,
the full spectra of an $8^4$ lattice and a $12^4$ lattice at
$\beta$=2.3 in SU(2) were obtained for one configuration.
In SU(2) one knows {\it a priori} that
every eigenvalue is doubly degenerate so that the number of single
positive eigenvalues must be $V/2$. Furthermore, they satisfy the
closure relation $\sum^{V/2}_{n=1} \lambda^2_n=V$. We found exactly
2048 single positive eigenvalues for the $8^4$ lattice and
10368 for the $12^4$ lattice, and the closure relations were
satisfied to a few parts in $10^8$.
This indicates that we can determine the spectrum accurately.

\subsection{Results for SU(2)}

A heat-bath Monte Carlo algorithm was used to generate quenched
gauge field configurations using the standard Wilson plaquette action,
with periodic boundary conditions on a $14^4$ lattice at $\beta$=2.2,
2.3, 2.4 and 2.5.
Anti-periodic boundary conditions were used for fermions in
all directions. Other boundary conditions were also tried,
and our results showed little change.
Gauge fixing was done with the help of
overrelaxation~\cite{man} which reduced the number of iterations by
a factor of 3 to 5. About 500 iterations with overrelaxation were
required for a stopping criterion $\delta \sim 10^{-6}$.

\begin{figure}
\psfig{file=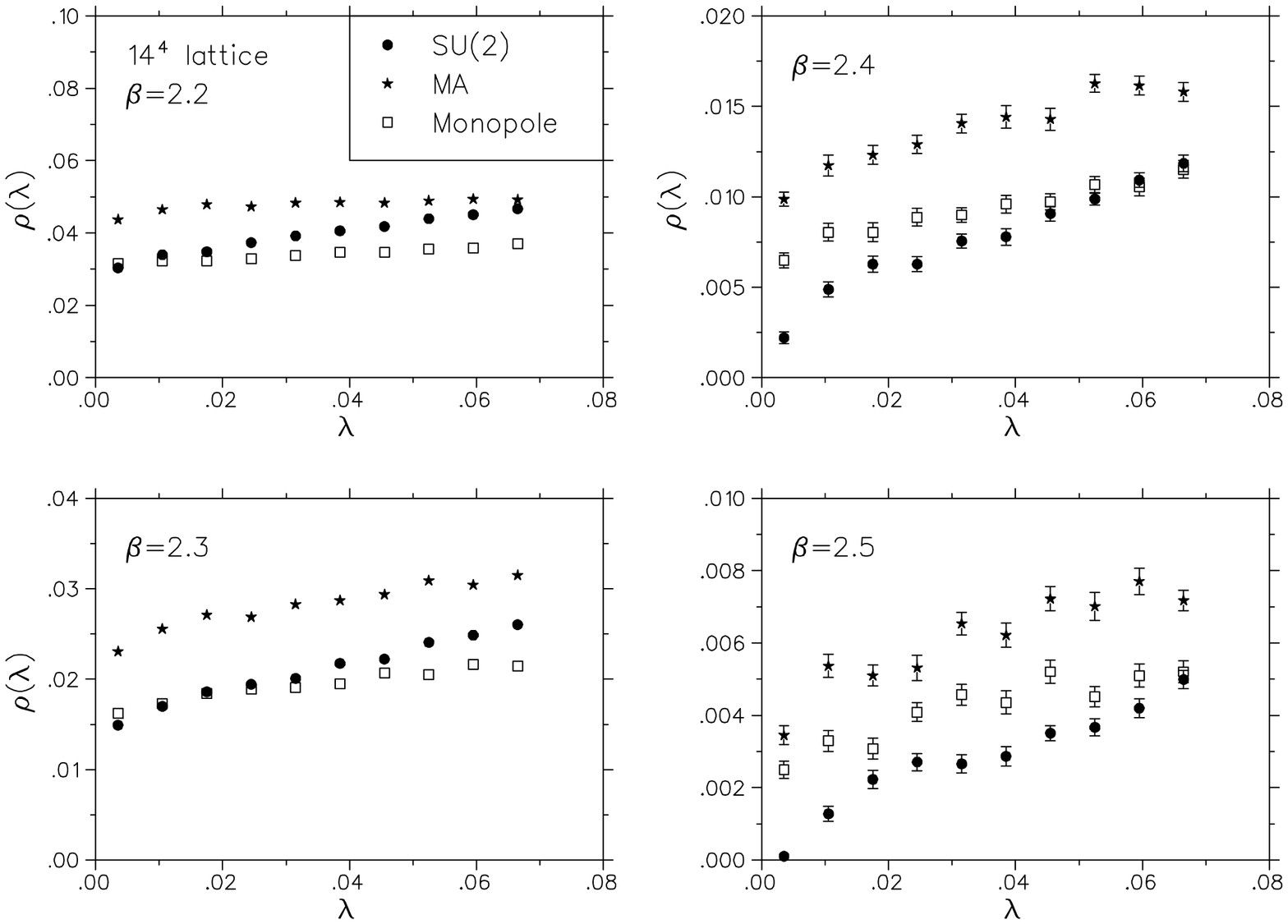,width=14cm,angle=90}
 \caption{Spectral density functions calculated at four different
$\beta$ values for three types of gauge configurations:
the non-Abelian SU(2) configurations, the Abelian-projected
configurations in the Maximal Abelian gauge, and the monopole
configurations further decomposed from the Abelian-projected
configurations.}
 \label{raw2}
 \end{figure}

\begin{figure}
\psfig{file=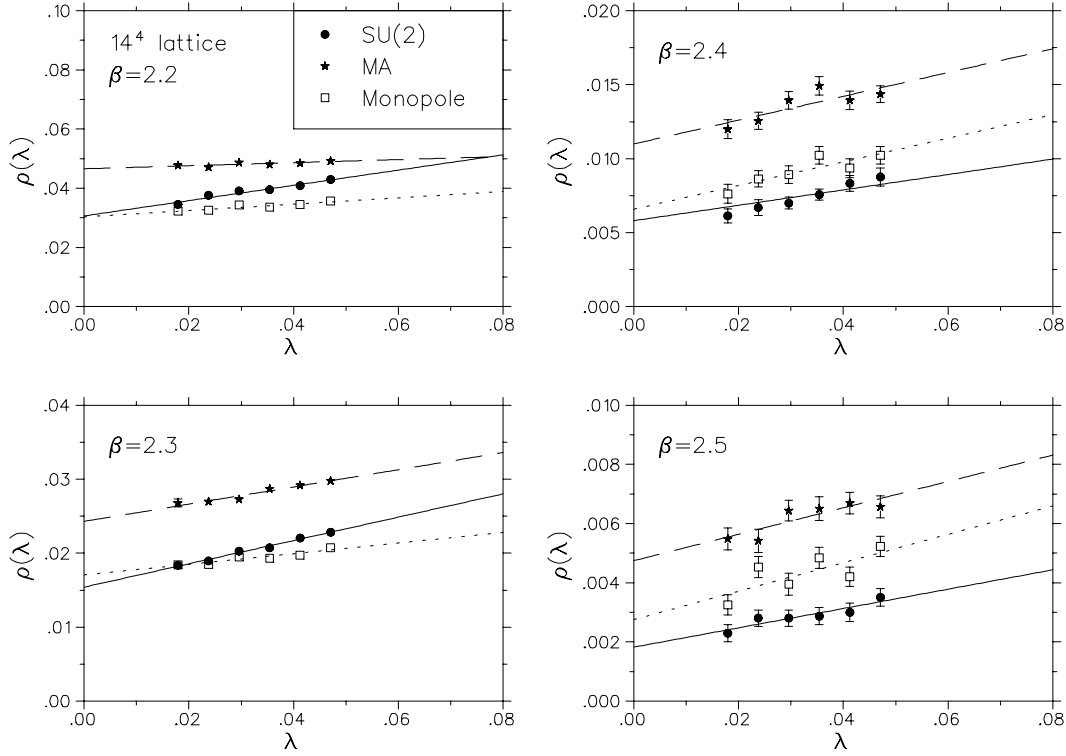,width=14cm,angle=90}
 \caption{Fitted spectral density functions in the interval
[0.015,0.05] in SU(2).}
 \label{fit2}
 \end{figure}

In Fig.~\ref{raw2} we show the raw data for the spectral density
function $\rho(\lambda)$ obtained from 70 configurations at each $\beta$
value. The results clearly show a non-vanishing signal
of $\rho(\lambda)$ at all $\beta$ values,
although finite volume effects manifested as a sudden depletion 
near $\lambda=0$ \cite{hands}, begin to set in at $\beta$=2.4 and 2.5.
To extract a value at $\lambda=0$, we fit the distributions by
a straight line $\rho(\lambda)=\rho(0)+\rho^\prime(0)\lambda$ 
in an interval $[\lambda_{min},\lambda_{max}]$.
The interval is chosen so that it excludes those eigenvalues
near $\lambda=0$ that are strongly influenced
by finite volume effects, and those that cause $\rho(\lambda)$ to
depart from linear behavior. It is expected that
$\lambda_{min}\sim \xi^3/V$ where $\xi$ is some
length scale governed by the gluon dynamics~\cite{hands}.
The value $\lambda_{max}$ is chosen to be as large as
possible while preserving both the stability  and
the quality of the fit. Under these conditions, we find that
the fit is quite stable over a relatively wide interval.
Fig.~\ref{fit2} shows the results of such a
fit for the eigenvalue interval [0.015,0.05].
The fitted values for $\rho(0)$ and $\rho^\prime(0)$ and the extracted
chiral condensate $\langle \bar{\chi} \chi \rangle$
are given in Table~\ref{chi2} in lattice units.
The errors quoted are statistical and are obtained using the
jackknife method. Since the eigenvalue spectrum is doubly degenerate,
the quoted numbers for $\rho(\lambda)$ are only half of the
true value, so that the extracted
$\langle \bar{\chi} \chi \rangle=-2\pi \rho(0)$.
The value of $\langle \bar{\chi} \chi \rangle$ in
the full SU(2) theory and in maximal Abelian gauge projection
are consistent with those obtained in Ref.~\cite{rmw}, where
an extrapolation to the chiral limit was made from nonvanishing
quark mass. The interesting feature here is that the monopole
configuration average has a condensate that is even closer
to the full theory than the Abelian projected fields.
On the other hand with the photon configurations
either no or very few small eigenvalues are found.
Hence we conclude there is no chiral symmetry breaking from these
configurations.

\begin{figure}
\psfig{file=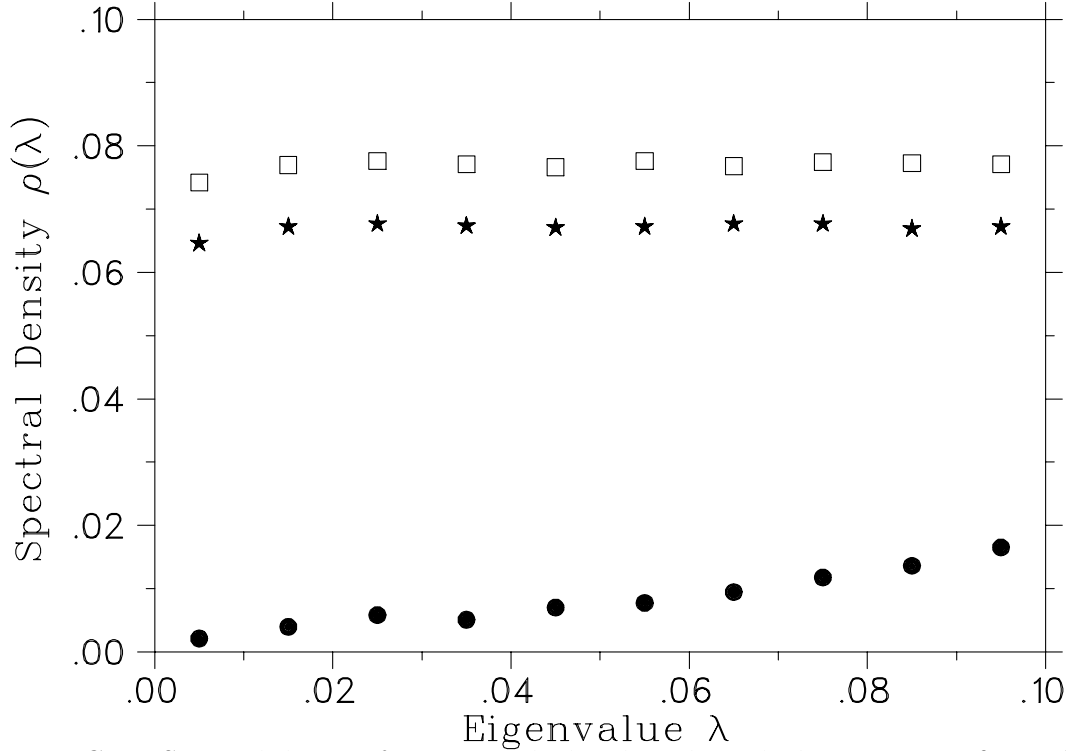,width=14cm,angle=90}
\caption{Spectral density functions 
calculated in the Polyakov gauge at $\beta$=2.5
for three types of gauge configurations:
non-Abelian SU(2)($\bullet$), Abelian($\Box$) and monopole($\star$).}
 \label{poly}
 \end{figure}

For purposes of comparison, we also performed Abelian projection
at $\beta=2.5$ using a different gauge-fixing condition: the
Polyakov gauge, in which the Polyakov loop is diagonalized
according to Eq.~(\ref{diag}). The result is shown in Fig.~\ref{poly}.
We see that the Abelian and the monopole field spectral density functions
are almost an order of magnitude larger than those of the full theory.
In Ref.~\cite{rmw}, a similar result for $\langle \bar{\chi} \chi \rangle$
is found using the field-strength gauge.

\subsection{Results for SU(3)}
\begin{figure}
\psfig{file=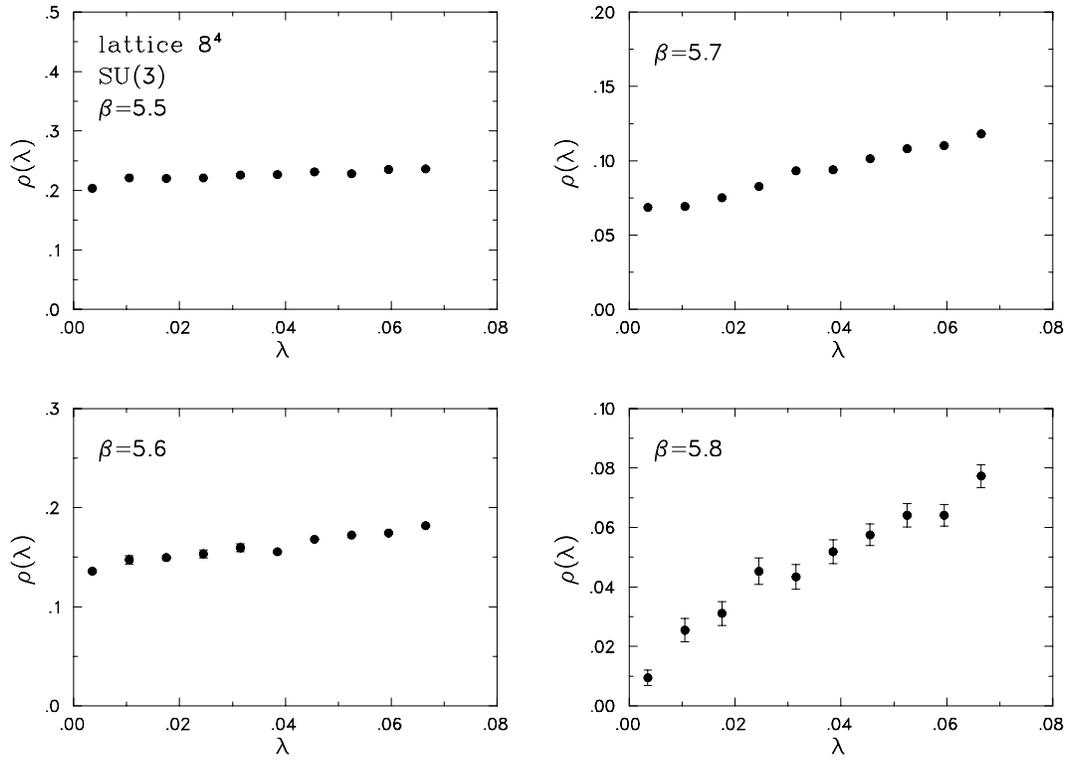,width=14cm,angle=90}
 \caption{Spectral density functions calculated at four different
$\beta$ values in SU(3).}
 \label{raw3}
 \end{figure}

The gauge field configurations were generated using the
Cabibbo-Marinari~\cite{cab} pseudo-heat-bath method on a
$8^4$ lattice at  $\beta$=5.5, 5.6, 5.7, 5.8 and
$10^4$ lattice at $\beta=5.9$.
Configurations are selected after 4000 thermalization sweeps from a
cold start, and every 500 sweeps thereafter.
Fig.~\ref{raw3} shows the spectral density function obtained from
44 SU(3) gauge field configurations for $\beta$=5.5,
66 configurations for $\beta$=5.6,
150 configurations for $\beta$=5.7,
and 37 configurations for $\beta$=5.8.

\begin{figure}
\psfig{file=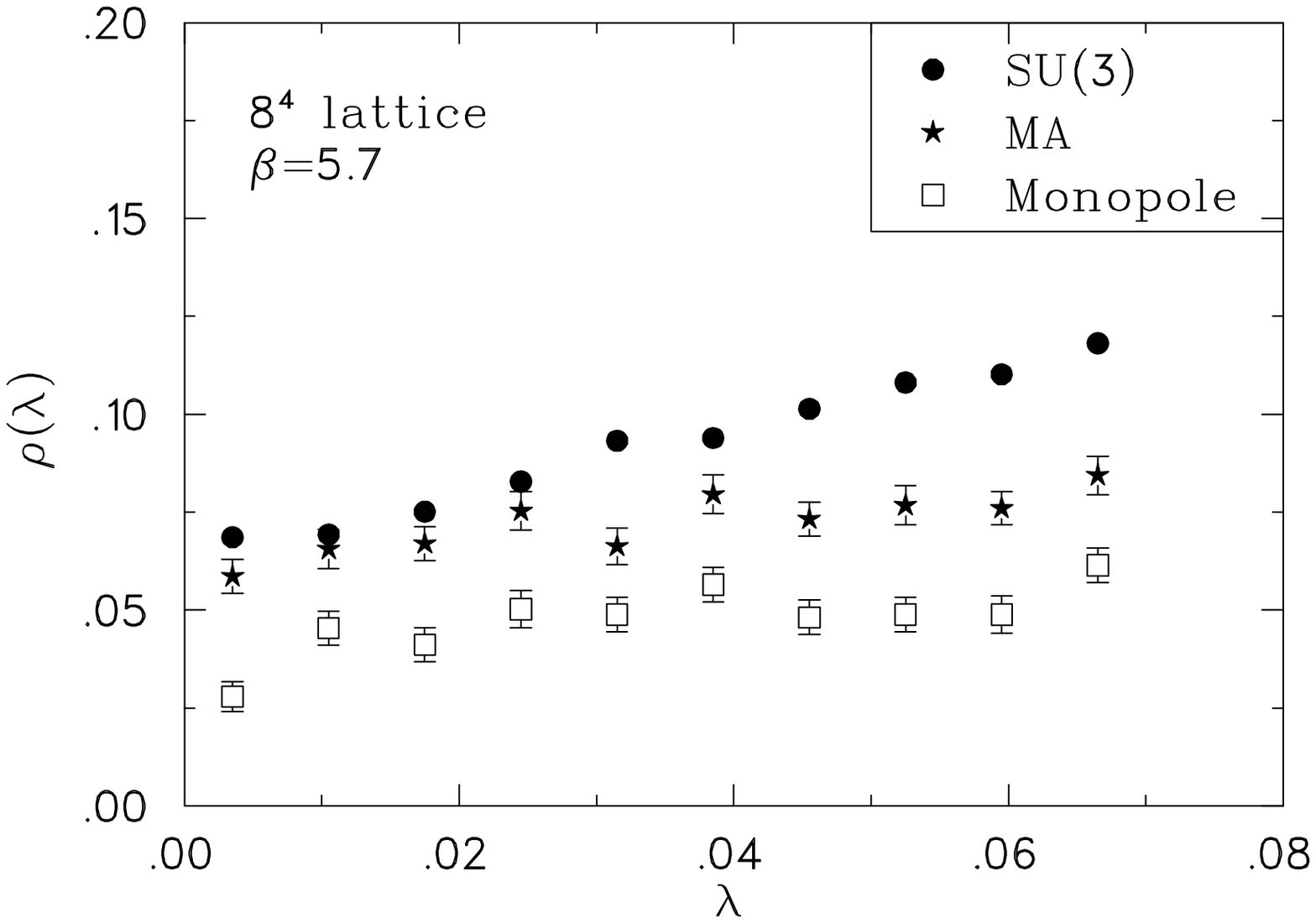,width=14cm,angle=90}
 \caption{Spectral density function in the SU(3) case on the
$8^4$ lattice at $\beta$=5.7.}
 \label{ch8b57}
 \end{figure}

\begin{figure}
\psfig{file=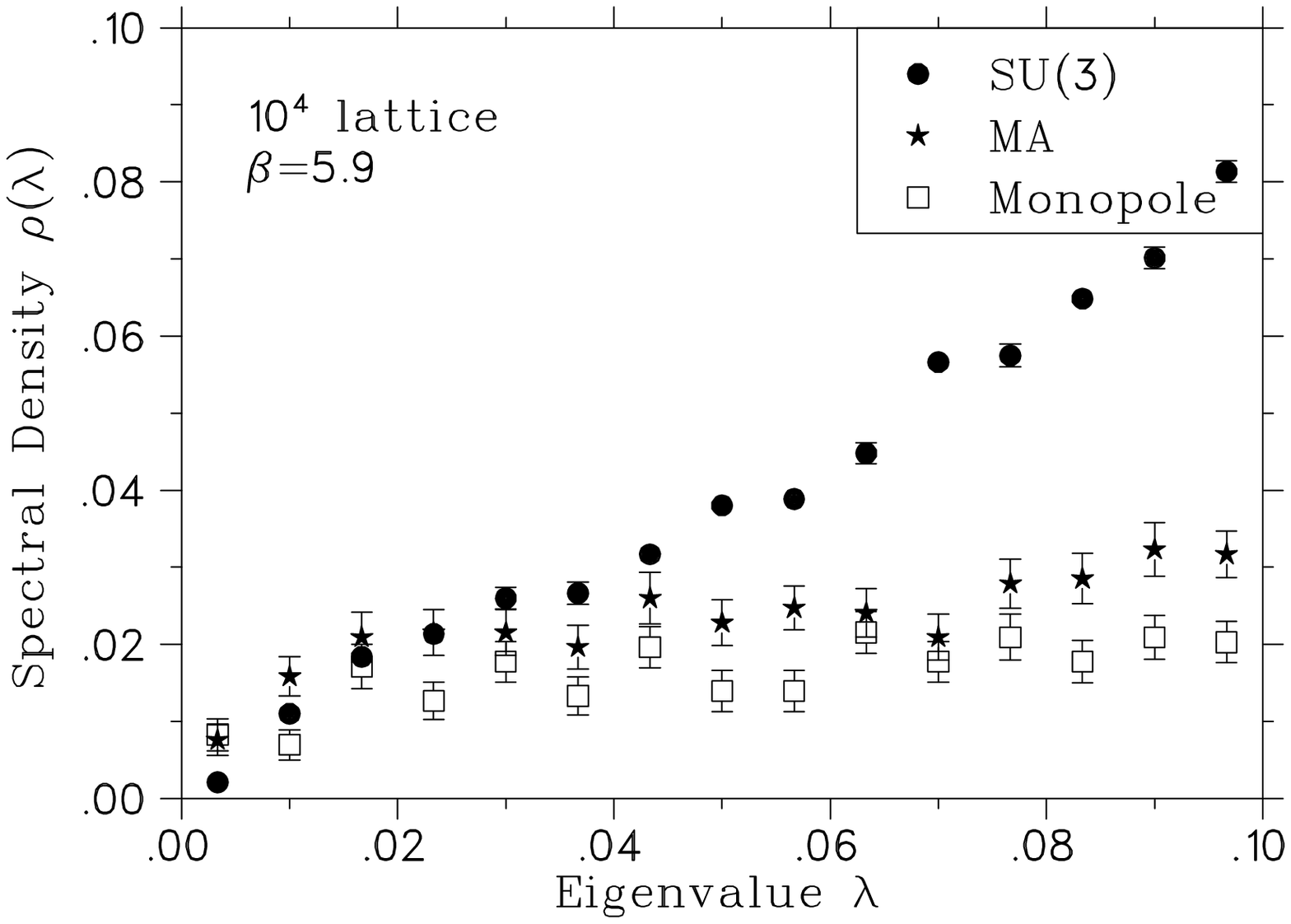,width=14cm,angle=90}
 \caption{Spectral density function in the SU(3) case on the
$10^4$ lattice at $\beta$=5.9.}
 \label{ch10b59}
 \end{figure}

We performed Abelian projections on the $8^4$ lattice
at $\beta=5.7$ and $10^4$ lattice at $\beta=5.9$.
Gauge fixing in SU(3) is time-consuming.
On average about 500 iterations with overrelaxation
are required for $\delta=1-1/9\mbox{Tr}[G_1(x)+G_2(x)+G_3(x)]$
to converge to $10^{-5}$.
After configuration averaging, the three phase factors 
in Eq.~(\ref{uuu}) give
equal contributions to the chiral condensate. So the total is
obtained by first calculating with one phase and then multiplying
by 3. This saves a factor of 3 in computer time.
Fig.~\ref{ch8b57} and Fig.~\ref{ch10b59} show the data obtained
for $\rho(\lambda)$.
For the $8^4$ lattice at $\beta=5.7$, 150 configurations were used,
and 100 configurations for $10^4$ at $\beta=5.9$.
We see that a similar pattern emerges in SU(3) as in SU(2):
for small eigenvalues, the Abelian and the monopole
contributions give spectral densities that are close
to those of the full theory.
It was also confirmed that photon configurations give
negligible contributions.

\section{Conclusion}
We have calculated chiral symmetry breaking on the lattice in
the quenched approximation and Abelian projection.
Unlike the previous study\cite{rmw}, the Lanczos method is used
to calculate eigenvalues of the fermion matrix and hence the
chiral condensate directly at zero quark mass. Furthermore, in this
work, a decomposition of the Abelian projected field into monopole and
photon pieces was made. For SU(2) gauge theory it was found the monopole
part of the Abelian field, projected in maximal Abelian gauge, yields
chiral condensate values which are quite close to those obtained with the
full non-Abelian fields. In contrast, the photon piece of the
Abelian projected field gives no condensate.

A calculation was also done using Abelian fields projected in the
Polyakov gauge. This yielded an eigenvalue density and hence a chiral
condensate about an order of magnitude larger than the non-Abelian
calculation. This is consistent with what was previously found in
field strength gauge in Ref.\cite{rmw}.

Some calculations were also done in an SU(3) theory. Qualitatively, a
similar pattern is seen as in the SU(2) calculation. In the region of
small eigenvalues (which is relevant for chiral symmetry breaking),
the Abelian and monopole fields give spectral
densities which are quite close to that of the full non-Abelian
calculation. This provides a positive indication that the idea
of Abelian dominance can be extended to the SU(3) theory.

\acknowledgments
This work was supported in part by the Natural Sciences and Engineering
Council of Canada.

 \begin{table}
\caption{The extracted values for $\langle \bar{\chi} \chi \rangle$
on the $14^4$ lattice from the straight line fit
$\rho(\lambda)=\rho(0)+\rho^\prime(0)\lambda$
in the interval [0.015,0.05].
At each $\beta$ value, the first row is for the full SU(2),
second row Maximal Abelian, third row monopole.
Numbers in parentheses are statistical errors in the final digit(s).}
 \label{chi2}
\begin{tabular}{cccc}
$\beta$ & $\rho(0)$ & $\rho^\prime(0)$ & 
${-\langle\bar{\chi}\chi\rangle}$\\ \hline
  2.2   & 0.0307(4) &  0.258(10) &  0.193(2) \\
        & 0.0466(5) &  0.052(14) &  0.293(3) \\
        & 0.0303(5) &  0.107(14) &  0.190(3) \\
  2.3   & 0.0154(4) &  0.158(11) &  0.097(3) \\
        & 0.0243(5) &  0.116(13) &  0.153(3) \\
        & 0.0170(5) &  0.072(13) &  0.107(3) \\
  2.4   & 0.0058(6) &  0.052(21) &  0.036(4) \\
        & 0.0110(7) &  0.080(19) &  0.069(4) \\
        & 0.0066(7) &  0.080(21) &  0.041(5) \\
  2.5   & 0.0018(3) &  0.033(09) &  0.011(2) \\
        & 0.0047(5) &  0.045(13) &  0.030(3) \\
        & 0.0028(4) &  0.048(12) &  0.017(3) \\
\end{tabular}
 \end{table}

\end{document}